\documentclass[aps,amsmath,amssymb,onecolumn]{revtex4}
\usepackage{graphicx}
\usepackage{dcolumn}
\usepackage{bm}
\usepackage{amsmath}
\usepackage{epstopdf}
\usepackage{amsfonts}
\usepackage{amssymb}
\usepackage{epsfig}
\usepackage{tabularx}
\usepackage{color}
\usepackage{soul}
\usepackage{multirow}
\usepackage{xcolor}

\usepackage[colorlinks=true,citecolor=blue,urlcolor=magenta,breaklinks]{hyperref}
\newcommand{\be}{\begin{equation}}
\newcommand{\en}{\end{equation}}
\newcommand{\bea}{\begin{eqnarray}}
\newcommand{\ena}{\end{eqnarray}}

\begin{document}

\title{Novel charged black hole solutions of Born-Infeld type: 
\\
General properties, Smarr formula and Quasinormal frequencies}

\author{
Leonardo Balart  {$^{a}$
\footnote{
\href{mailto:leonardo.balart@ufrontera.cl}{leonardo.balart@ufrontera.cl} 
}
}
Sebasti{\'a}n Belmar-Herrera {${}^{a}$
\footnote{
\href{mailto:sebastian.belmar@ufrontera.cl}{sebastian.belmar@ufrontera.cl} 
}
} 
Grigoris Panotopoulos  {${}^{a}$
\footnote{
\href{mailto:grigorios.panotopoulos@ufrontera.cl}{grigorios.panotopoulos@ufrontera.cl} 
}
}
Angel Rincon  {${}^{b}$
\footnote{
\href{mailto:angel.rincon@ua.es}{angel.rincon@ua.es} 
}
}
}

\address{
${}^a$ Departamento de Ciencias F{\'i}sicas, Universidad de la Frontera, Casilla 54-D, 4811186 Temuco, Chile.
\\
${}^b$ Departamento de Física Aplicada, Universidad de Alicante,
Campus de San Vicente del Raspeig, E-03690 Alicante, Spain.
}

\begin{abstract}
We investigate two novel models of charged black holes in the framework of non-linear electrodynamics of Born-Infeld type. In particular, starting from two concrete Lagrangian densities, the corresponding metric potentials, the electric field, the Smarr formula and finally, the (scalar) quasinormal modes are computed for each model. Our findings show that although the models look very similar, their quasinormal spectra are characterized by certain differences.
\end{abstract}

\maketitle

\section{Introduction}
\label{intro}

Non-linear Electrodynamics (NLE), with already a long history, has attracted a lot of attention, and it has been extensively studied over the years. To begin with, one may mention that classical electrodynamics, i.e. Maxwell's theory, consists of a system of linear equations. At quantum level, however, as soon as radiative corrections are taken into account, the effective field equations become non-linear. The very first works go back to the 30's, when Euler and Heisenberg managed to obtain corrections within quantum electrodynamics \cite{Euler}, while Born and Infeld were able to obtain a finite self-energy of point-like charges \cite{BI}. Born-Infeld theory has inspired other models of NLE, for example the theory of Soleng \cite{soleng} and others, such as those presented in Refs. \cite{Kruglov:2014hpa, Kruglov:2016cdm, Kruglov:2017fuj, Kruglov:2017xmb, Gaete:2017cpc, Gaete:2018nwq, Mazharimousavi:2019sgz, Mazharimousavi:2021uki, Gullu:2020ant,Bandos:2020jsw, Kruglov:2021bhs, Belmar-Herrera:2022ebd}. They all characterized by a non-linear parameter $b$, and in a certain limit ($b \rightarrow 0$
or $b \rightarrow \infty$, depending on the formulation) the standard Reissner-Nordstr{\"o}m solution \cite{RN} 
of Maxwell's linear theory is recovered. 
Thus, as already mentioned, Born and Infeld introduced non-linear electrodynamics to avoid divergences obtaining a finite self-energy of a point-like charge. For two decades, the Born-Infeld action was extensively studied  in the context of Superstring Theory, where the dynamics of D-branes, solitonic solutions of Supergravity, are described by the corresponding Born-Infeld action. In this sense, the replacement of the well-known Reissner-Nordstrom black hole solutions (in Einstein-Maxwell theory) to the charged black hole solutions (in Einstein-Born-Infeld theory) has motivated several investigations, for example, see \cite{Aiello:2004rz,Dey:2004yt}. More importantly, albeit we have a vast list of non-linear electrodynamics black hole solutions, many of them still maintain the singularity problem and just offer a more intricate solution, different from the Born-Infeld black hole solution where a non-divergent electric field is obtained.
Additional works on non-linear electrodynamics and their properties can be found in \cite{Kanzi:2020cyv,Jafarzade:2020ova,Pourhassan:2022cvn,deOliveira:1994in,Javed:2022psa,Kumaran:2022soh,Okyay:2021nnh,Javed:2019kon}.

\smallskip 

Moreover, a generalization of Maxwell's theory leads to the so called Einstein-power-Maxwell (EpM) theory \cite{EpM1,EpM2,EpM3,EpM4,EpM5,EpM6,EpM7,EpM8,EpM9,EpM10,EpM11,Hassaine:2008pw,Gonzalez:2009nn}, described by a power-law where the Lagrangian density is of the form $\mathcal{L}(F) = F^k$, with $F$ being the Maxwell invariant, and where $k$ is an arbitrary rational number. The advantage of EpM theories is that they preserve the nice properties of conformal invariance in any number of space-time dimensionality $D$, as long as $k=D/4$, since it is easy to verify that for that choice of $k$ the corresponding stress-energy tensor is trace-less. Furthermore, considering appropriate non-linear sources, which in the weak field limit boil down to Maxwell's theory, one may generate new solutions (Bardeen-like solutions \cite{Bardeen}, see also \cite{borde}) to Einstein's field equations \cite{beato1,beato2,beato3,bronnikov,dymnikova,hayward,vagenas1,vagenas2,Rodrigues:2017yry}. Those new solutions possess an event horizon, while at the same time their curvature invariants are regular everywhere. This is to be contrasted to the standard Reissner-Nordstr{\"o}m solution of Einstein-Maxwell theory, where a singularity at the origin, $r \rightarrow 0$, is present.

\smallskip

Black holes are a robust prediction of Einstein's General Relativity \cite{GR}. They are fascinating objects of paramount
importance both for classical and quantum gravity. Over the last years, after the first image of
the black hole shadow by the Event Horizon Telescope \cite{EHT}, together with the historical first direct detection of
gravitational waves from a black hole merger in binaries by the LIGO/VIRGO collaborations \cite{GWs}, we have been convinced 
that black holes do exist in Nature. Static, spherically symmetric black hole solutions with a net electric charge 
$Q$ and mass $M$ within linear or non-linear electrodynamics have thermal properties, just like ordinary thermodynamic systems, and they obey the four laws of black hole thermodynamics \cite{Bardeen:1973gs}, which are very similar to the usual four 
laws of statistical mechanics. In particular, black hole solutions have a Hawking temperature, $T_H$, and a Bekenstein entropy, $S$, given by \cite{hawking1,hawking2,bekenstein}
\begin{equation}
T_H=\frac{1}{4 \pi} \: f'(r_h), \; \; \; \; S=\frac{A_h}{4}=\pi r_h^2
\end{equation}
where $r_h$ is the event horizon, $f(r)$ is the lapse function, $r$ is the radial coordinate, and $A_h=4 \pi r_h^2$ 
is the horizon area for solutions with a spherical horizon topology. Moreover, one may derive a Smarr formula \cite{smarr} 
using Euler's theorem for homogeneous functions. Within Maxwell's linear electrodynamics, the expression for the Hawking 
temperature is the same irrespectively of how it is calculated, namely using any of the following equations
\begin{equation}
T_H=\frac{1}{4 \pi} \: f'(r_h), \; \; \; \; T_H=\left( \frac{\partial M}{\partial S} \right)_Q
\end{equation}
while the Smarr formula 
\begin{equation}
M=2 T_H S+\Phi_h Q
\end{equation}
is compatible with the first law of black hole mechanics
\begin{equation}
dM = T_H dS + \Phi_h dQ
\end{equation}
where $\Phi_h$ is the electric potential, $\Phi(r)$, evaluated at the horizon, $\Phi_h \equiv \Phi(r=r_h)$. However,
for black hole solutions in NLE, it is not always possible to obtain a Smarr formula compatible with the first law of
black hole mechanics (see for example \cite{Balart:2017dzt}). In addition to that, the two ways to calculate Hawking temperature lead to two different expressions \cite{Ma:2014qma}.

\smallskip

Realistic black holes are not isolated in Nature. Instead, they are in constant interaction with their environment.
When a black hole is perturbed, the geometry of space-time undergoes dumped oscillations. How a system responds to 
small perturbations has always been important in Physics. The work of \cite{regge} marked the birth of
black hole perturbations, it was later extended by \cite{zerilli1,zerilli2,zerilli3,moncrief,teukolsky}, while the state-of-the art in black hole perturbations is 
summarized in the comprehensive review of Chandrasekhar's monograph \cite{monograph}. Quasi-normal (QN) frequencies, with a non-vanishing 
imaginary part, are complex numbers that encode the information on how black holes relax after the perturbation has been 
applied. They depend on the geometry itself as well as the type of the perturbation (scalar, Dirac, vector (electromagnetic), tensor (gravitational)). As 
they do not depend on the initial conditions, QN modes (QNMs) carry unique information about black holes. Black hole perturbation
theory and QNMs of black holes are relevant during the ringdown phase of binaries, in which after the merging of two black 
holes a new, distorted object is formed, while at the same time the geometry of space-time undergoes dumped 
oscillations due to the emission of gravitational waves.

\smallskip

Given the interest in gravitational wave Astronomy and in QNMs of black holes, it would be interesting to see what kind
of QN spectra are expected from electrically charged black holes in non-linear electrodynamics. In the present work we 
propose to compute QNMs for scalar perturbations of black hole solutions with a net electric charge
within two NLE models obtained recently starting from thermodynamic properties. The solutions of each model exhibit 
three distint behaviors (RN type, Schwarzschild type or marginal) depending on the values of the parameters.

\smallskip

Our work is organized as follows: After this introduction, in the next section we present the models of NLE. In Section 3 we discuss charged black hole solutions and their basic properties, and then in the fourth section we analyze their thermodynamics. The propagation of massless scalar fields into a spherically symmetric gravitational background as well as the WKB method of sixth order are described in Section 5, while in the sixth Section we present and discuss our main results. Finally, we conclude in Section 7 summarizing our work. We adopt the metric 
signature $-,+,+,+$, and we work in geometrical units where we set the speed of light in vacuum and Newton's constant 
to unity, $G=1=c$.



\section{Non-linear electrodynamics models}
\label{NLE}

In this section we present two nonlinear electrodynamics models, which are characterized in that, as in the Born-Infeld model, the self-energy for charged particles is regularised and also depend on a nonlinearity parameter $b$ that in both cases is associated with the maximum value of the electric field. Both models are analytically tractable
and, unlike others, the expansion in powers of the Maxwell invariant $\mathcal{F}$ produces a series with non-polynomial powers.
The first of them is presented in this work and the other was presented in Ref.~\cite{Mazharimousavi:2019sgz}.

\smallskip

The first model is represented by the following Lagrangian
\begin{equation}
\mathcal{L}(\mathcal{F}) = \frac{- 4 b \mathcal{F}}{\left(\sqrt{b}+ \sqrt{b-2\sqrt{-\frac{\mathcal{F}}{2}}} \right)^2}\, .
\label{Lag-1}
\end{equation}
Here $\mathcal{F} = F_{ \mu \nu} F^{ \mu \nu}/4 = (B^2 - E^2)/2$. Note that Maxwell electrodynamics is recovered by letting $b \rightarrow \infty$
\begin{equation}
\mathcal{L}(\mathcal{F}) = -\mathcal{F}-\frac{\sqrt{-\mathcal{F}} \mathcal{F}}{\sqrt{2} b}+\frac{5 \mathcal{F}^2}{8 b^2}+\frac{7 \sqrt{-\mathcal{F}} \mathcal{F}^2}{8
   \sqrt{2} \, b^3}+\mathcal{O}\left(\frac{1}{b^{4}}\right) \, .
\label{}
\end{equation}

Considering $B = 0$ and spherical symmetry for the electric field, then we can rewrite the electromagnetic field equations $\nabla_{\mu} \left( F^{\mu \nu} \mathcal{L}_{\mathcal{ F}}\right) = 0$ as follows 
\begin{equation}
\frac{1 }{r^2}\frac{d}{dr}[r^2 E(r)\mathcal{L}_{\mathcal{F}}] = 0
\,\,\label{field-eqs} \,  ,
\end{equation} 
where $\mathcal{L}_{\mathcal{F}}$ represents the derivative of $ \mathcal{L}(\mathcal{F})$ with respect to the Maxwell invariant given by $\mathcal{F} = - E(r)/2$.
If we replace the Lagrangian~(\ref{Lag-1}) in Eq.~(\ref{field-eqs}), and we solve the equation that results for $E(r)$, then we obtain
\begin{equation}
E(r) = \frac{b q \left(2 b r^2+q\right)}{2 \left(b r^2+q\right)^2} \, ,
\label{EF-1}
\end{equation}
which asymptotically behaves like a Coulombian field
\begin{equation}
E(r) = \frac{q}{r^2} -\frac{3 q^2}{2 b r^4} + \mathcal{O}\left( \frac{1}{r^6}\right)       \, .
\label{}
\end{equation}
In the limit $r \rightarrow 0$ the electric field behaves as
\begin{equation}
E(r) = \frac{b}{2} -\frac{b^3 r^4}{2 q^2} +\mathcal{O}\left(r^6 \right)     \, .
\label{}
\end{equation}
The maximum value of the electric field is obtained at $E(0)$.

\smallskip

The Lagrangian of the other model is written as~\cite{Mazharimousavi:2021uki}
\begin{equation}
\mathcal{L}(\mathcal{F})=  -\frac{\sqrt{2} b \mathcal{F}}{\sqrt{2} b - 2 \sqrt{- \mathcal{F}}} \, .
\label{Lag-2}  
\end{equation}
Considering the limit of very large $b$, we obtain
\begin{equation}
\mathcal{L}(\mathcal{F}) = -\mathcal{F}-\frac{\sqrt{-\mathcal{F}} \mathcal{F}}{b}+\frac{\mathcal{F}^2}{b^2}+\frac{\sqrt{-\mathcal{F}}\mathcal{F}^2}{b^3} + \mathcal{O}\left(\frac{1}{b^4}\right) \, .
\label{}  
\end{equation}

In this model, the electric field obtained is
\begin{equation}
E(r) = b \left(1-r \sqrt{\frac{b}{b r^2+2 q}}\right) \, .
\label{EF-2}  
\end{equation}
In the limit $r \rightarrow \infty$ it behaves like a Coulombic field. Furthermore the electric field is maximum when $E(0) = b$.

\smallskip

\begin{figure}[h]
\centering
\includegraphics[scale=0.65]{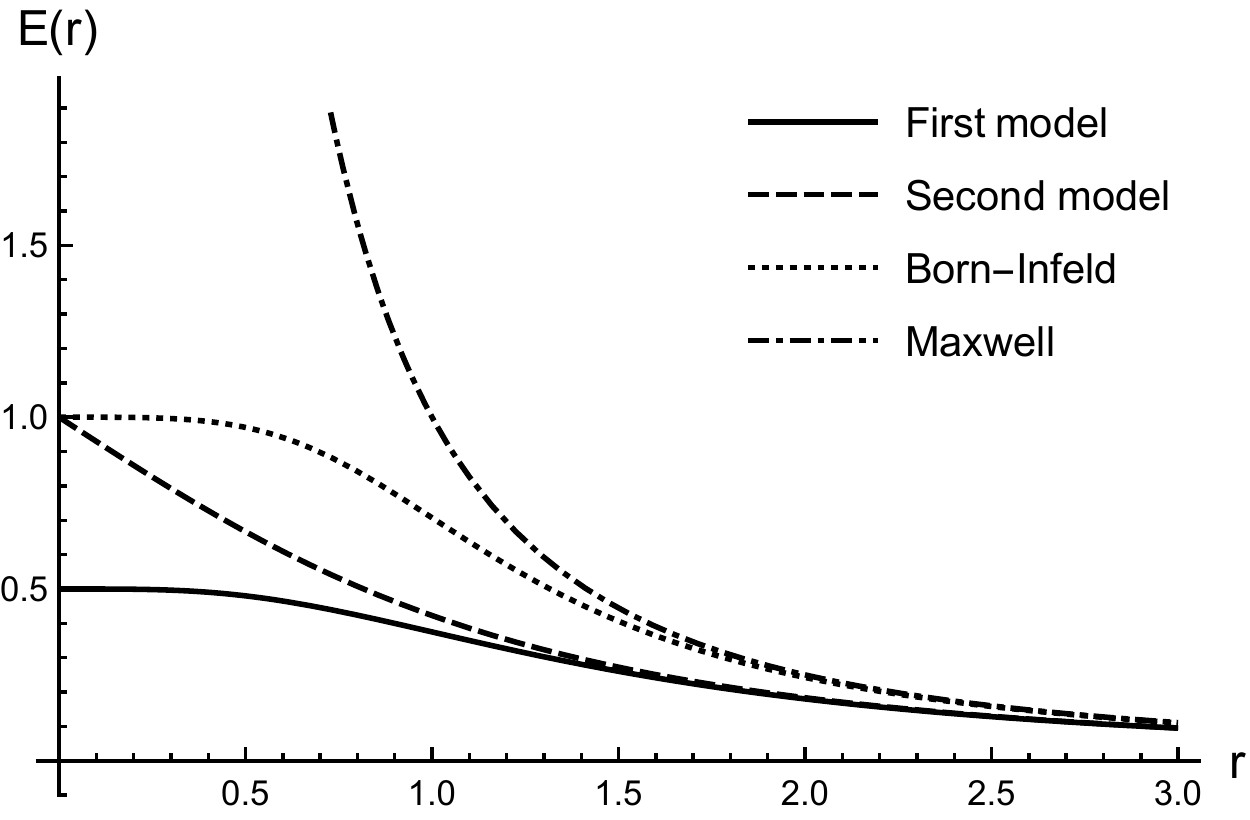}  
\caption{Electric fields of the two models studied together with those corresponding to the Maxwell and Born-Infeld models. Here $b=1$ and $q=1$ in all cases.}
\label{Fig-E-1-2}
\end{figure}

Figure~\ref{Fig-E-1-2} shows the electric fields as a function of $r$ that are obtained for both models. Also included are the Maxwell and Born-Infeld electric fields, respectively, with the same values for the parameter $b$ and the electric charge $q$.

\smallskip

As the authors point out in Ref.~\cite{Mazharimousavi:2021uki}, the model represented by the Lagrangian~(\ref{Lag-2}) is the electrical counterpart of the magnetic solution given in Ref.~\cite{Kruglov:2017xmb}. Let us add that the magnetic counterpart of the Lagrangian~(\ref{Lag-1}) can be obtained in the same way.


\section{Charged black hole solutions within nonlinear electrodynamics: Novel models}
\label{BH}

The action of Einstein's General Relativity coupled to any of the nonlinear electrodynamics models from the previous section is represented by
\begin{equation}
S =   \int d^4x \sqrt{-g} \left[ \frac{R - 2 \Lambda}{16\pi G} + \mathcal{L}(\mathcal{F})\right]    \, ,
\label{}
\end{equation}
where $g$ is the determinant of the metric tensor, $R$ is the Ricci scalar, $\Lambda = -3/l^2$ is the cosmological constant and $\mathcal{L}(\mathcal{F})$ is the Lagrangian given by Eq.~(\ref{Lag-1}) or that of Eq.~(\ref{Lag-2}).

\smallskip

The variation of the action yields the fields equations
\begin{equation}
G_{\mu \nu} + \Lambda g_{\mu \nu} = R_{\mu \nu}-\frac{1}{2} R g_{\mu \nu} + \Lambda g_{\mu \nu} = 8 \pi T_{\mu \nu} \, ,
\label{field-eq}
\end{equation}
where the energy-momentum tensor $T_{\mu \nu}$ is given by
\begin{equation}
T_{\mu \nu}=\frac{1}{4 \pi} \left(g_{\mu \nu} \mathcal{L} -  F_{\mu \alpha} F_{\nu}^{\, \alpha}  \mathcal{\mathcal{L}}_{\mathcal{F}} \right)  \, ,
\label{}
\end{equation}
and 
\begin{equation}
\nabla_{\mu} \left( F^{\mu \nu} \mathcal{L}_{\mathcal{F}}\right)=0 \, .
\label{}
\end{equation}

\smallskip

We consider static spherically symmetric black hole solutions described by the metric 
\begin{equation}
ds^2 = -f(r) dt^2 + f^{-1}(r) dr^2+ r^2 (d\theta^2 + \sin^2 \theta d\phi^2)  \, .
\label{metric}
\end{equation}


\subsection{First model}
\label{1model}

If we consider the metric~(\ref{metric}) in the fields equations~(\ref{field-eq}), then from the component $(tt)$ or $(rr)$ we get
\begin{equation}
\frac{-1+ f(r) + r f^{\prime}(r) }{r^2} + \Lambda = 2 \mathcal{L} + 2 E(r)^2 \mathcal{\mathcal{L}}_{\mathcal{F}}
\,\,\label{} \,  
\end{equation}
and from component $(\theta\theta)$ or ($\phi\phi)$
\begin{equation}
\frac{f^{\prime}(r) }{r} +  \frac{f^{\prime\prime}(r) }{2}+ \Lambda = 2 \mathcal{L}
\,\,\label{} \,  .
\end{equation} 
Carrying out the subtraction between these last two expressions results in
\begin{equation}
\frac{-1+ f(r) }{r^2} -\frac{f^{\prime\prime}(r) }{2}+ \Lambda = 2 E(r)^2 \mathcal{L}_{\mathcal{F}}
\,\,\label{} \,  .
\end{equation} 
If we now take $\mathcal{L}$ given by Eq.~(\ref{Lag-1}) which depends on $\mathcal{F} = -E(r)^2/2$, where $E(r)$ is given by expression~(\ref{EF-1}), we obtain a differential equation whose solution for $f(r)$ is
\begin{align}
\begin{split}
f(r) =  1 - &\frac{2 M}{r} - \frac{\Lambda r^2}{3} + \frac{\pi  \sqrt{b q^3}}{2 r}   - 
\frac{\sqrt{b} q^{3/2} }{r}  \tan^{-1}\left(\frac{\sqrt{b} \, r}{\sqrt{q}}\right)  \, ,
\label{mf-1}
\end{split}
\end{align}
where is the mass of the black hole, $q$ the electric charge and $\Lambda < 0$. When the parameter $b \rightarrow \infty$
we recover the Reissner-Nordstr{\"o}m AdS solution.

\smallskip

The black hole that we have presented, as in the case of the Born-Infeld black hole~\cite{Fernando:2003tz, Fernando:2006gh, Gunasekaran:2012dq, Breton:2017hwe}, has one or two horizons depending on the value of the parameters $M$, $q$ and $b$. To classify the solutions we analyze the behavior of the metric function in the limit $r \rightarrow 0$
\begin{align}
\begin{split}
f (r) = 1  & - \frac{2M - \frac{1}{2} \pi  \sqrt{b q^3}}{r} - b q -  \frac{\Lambda r^2}{3} + \frac{b^2
   r^2}{3}
   -\frac{b^3 r^4}{5 q} + \mathcal{O}\left(r^6\right) \, ,
\end{split}
\end{align}
We can rewrite this equation as
\begin{equation}
f(r) = 1-\frac{2(M - a)}{r}- b q + \mathcal{O}(r)  \,  ,
\label{serie_nuevo}
\end{equation} 
where $a$ (called marginal mass in Ref.~\cite{Gunasekaran:2012dq}) is
\begin{equation}
a = \frac{1}{4} \pi  \sqrt{b q^3} \, .
\end{equation}  

As was done with the Born-Infeld black hole~\cite{Fernando:2003tz, Fernando:2006gh, Gunasekaran:2012dq, Breton:2017hwe}, we can categorize the solutions that arise, depending on the mass $M$, the electric charge $q$ and the parameter $b$ (in our analysis we choose $\Lambda = 0$).
Here too there are three possible categories: Schwarzschild (S) type solution, which occurs when $M>a$ and has one event horizon. The marginal solution, present when $M=a$. And the Reissner-Nordström (RN) type solution, which occurs when $M<a$. 
Considering the expansion~(\ref{Lag-1}) for the marginal case
\begin{equation}
f(r) =1- b q+\mathcal{O}(r)   \, ,
\label{}
\end{equation} 
it follows that the solution has one event horizon if $b q > 1$ and presents a naked singularity if $b q < 1$.

\smallskip

To know the number of event horizons of the RN type solutions, we must calculate the extreme mass $M_{ex} \equiv M(r_{ex})$, where $r_{ex}$ is obtained from the expression
$\partial_r M(r)|_{r=r_{ex}} = 0$. In this case it is given by


\begin{align}
\begin{split}    
M_{ex} = & \frac{b q^{3/2}}{4 \sqrt{b}}  \left(\pi -2 \tan^{-1}\left(\sqrt{b q-1}\right)\right) +
\frac{2}{4 \sqrt{b}} \sqrt{q (b q-1)}\, .
\label{}
\end{split}
\end{align}

\begin{figure}[h]
\centering
\includegraphics[scale=0.90]{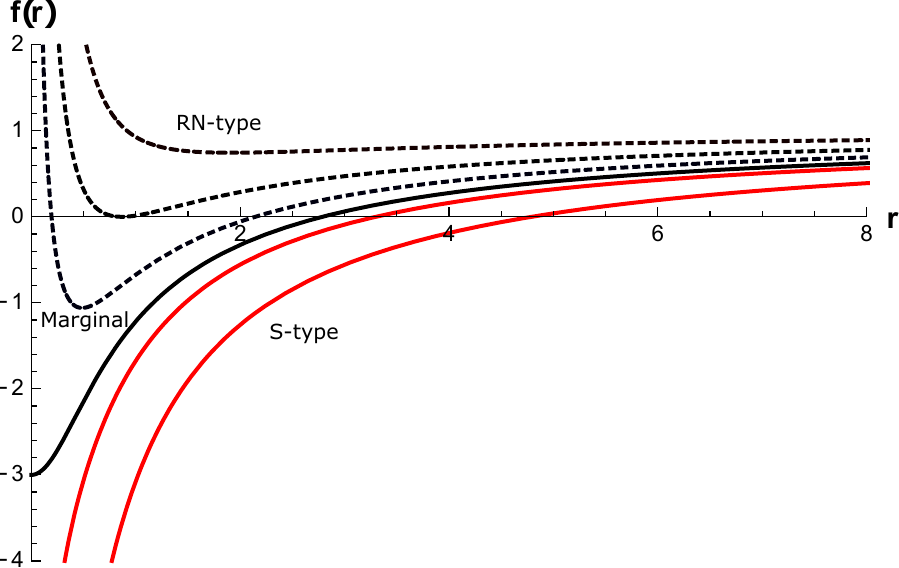}  
\caption{Different types of solutions corresponding to the metric function~(\ref{mf-1}) for different values of $M$. Here $b=4$, $q=1$ and $\Lambda = 0$ are used.}
\label{Fig-mf-1}
\end{figure}

 The types of possible solutions can be summarized according to the values of $M$, $b$ and $q$, in Table~(\ref{tab: lagr-1}). Figure~(\ref{Fig-mf-1}) illustrates the different types of solutions that can be classified for the metric function given by Eq.~(\ref{mf-1}). Note that the marginal case illustrated in this figure corresponds to a one of a regular metric function at the origin.

\begin{table}[h]
\centering
\begin{tabular}{cccc|}                                                                                                                                                                  \\ \hline
\multicolumn{2}{|c|}{Conditions}                                                                                                                                       & \multicolumn{1}{c|}{Type}     &  Horizons\\ \hline
\multicolumn{2}{|c|}{$M>a$}                                                                                                                               & \multicolumn{1}{c|}{S}        &    one           \\ \hline
\multicolumn{1}{|c|}{\multirow{3}{*}{$M<a$}} & \multicolumn{1}{c|}{\begin{tabular}[c]{@{}c@{}}$M<M_{ex}$  \\ $b q > 1$\end{tabular}}  & \multicolumn{1}{c|}{RN}       & zero          \\ \cline{2-4} 
\multicolumn{1}{|c|}{}                                  & \multicolumn{1}{c|}{\begin{tabular}[c]{@{}c@{}}$M>M_{ex}$\\ $b q > 1$\end{tabular}} & \multicolumn{1}{c|}{RN}       & two          \\ \cline{2-4} 
\multicolumn{1}{|c|}{}                                  & \multicolumn{1}{c|}{\begin{tabular}[c]{@{}c@{}}$M=M_{ex}$\\ $b q > 1$\end{tabular}}              & \multicolumn{1}{c|}{RN}       & one          \\ \hline
\multicolumn{1}{|c|}{\multirow{2}{*}{$M=a$}}           & \multicolumn{1}{c|}{$b q > 1$}                                                                     & \multicolumn{1}{c|}{Marginal} & one           \\ \cline{2-4} 
\multicolumn{1}{|c|}{}                                  & \multicolumn{1}{c|}{$b q < 1$}                                                                        & \multicolumn{1}{c|}{Marginal} & zero         \\ \hline
\end{tabular}
\caption{Classification of the types of solutions that can be obtained from the metric function~(\ref{mf-1}) as a function of the mass $M$.}
\label{tab: lagr-1}
\end{table}


\subsection{Second model}
\label{2 model}

Considering now the metric~(\ref{metric}) in the fields equations~(\ref{field-eq}) for the Lagrangian~(\ref{Lag-2}) and going through the same steps as in the previous subsection, one can obtain the corresponding metric function
\begin{align}
\begin{split}
f(r) = & 1-2 b q-\frac{2 M}{r} - \frac{\Lambda r^2}{3}  -\frac{2}{3} b^2 r^2 +
\frac{4 \sqrt{b} q \sqrt{b r^2+2 q}}{3 r} + 
\frac{2}{3} b^{3/2} r \sqrt{b r^2+2 q}  \, ,
\label{mf-2}
\end{split}
\end{align}
which behaves asymptotically as the Reissner-Nordstr{\"o}m solution.

\smallskip

The solution of this black hole, as in the previous case, has one or two horizons depending on the value of the parameters $M$, $q$ and $b$. To classify the types of solutions again, we can analyze the behavior of the metric function in the limit $r \rightarrow 0$, where
\begin{equation}
f(r) = 1-\frac{2(M - a)}{r}- 2 b q+\mathcal{O}(r)   \, .
\label{}
\end{equation} 
and 
\begin{equation}
a = \frac{2}{3} \sqrt{2b}  q^{3/2}   \, .
\label{a2}
\end{equation}  
As above we can also find the extremal mass
\begin{align}
\begin{split}
M_{ex} = &\frac{1}{24} \Biggr[  b q^2 \sqrt{(2 b q+1)^2}+4 q \sqrt{(2 b q+1)^2} - 
8 b^2 q^3 - 12 b q^2+18  q+\frac{1}{b} \sqrt{(2 b q+1)^2}-\frac{5}{b} \Biggr]\, .
\label{Mex2}
\end{split}
\end{align}
The analysis is similar if we consider the cosmological constant.

\smallskip

\begin{figure}[h!]
\centering
\includegraphics[scale=0.9]{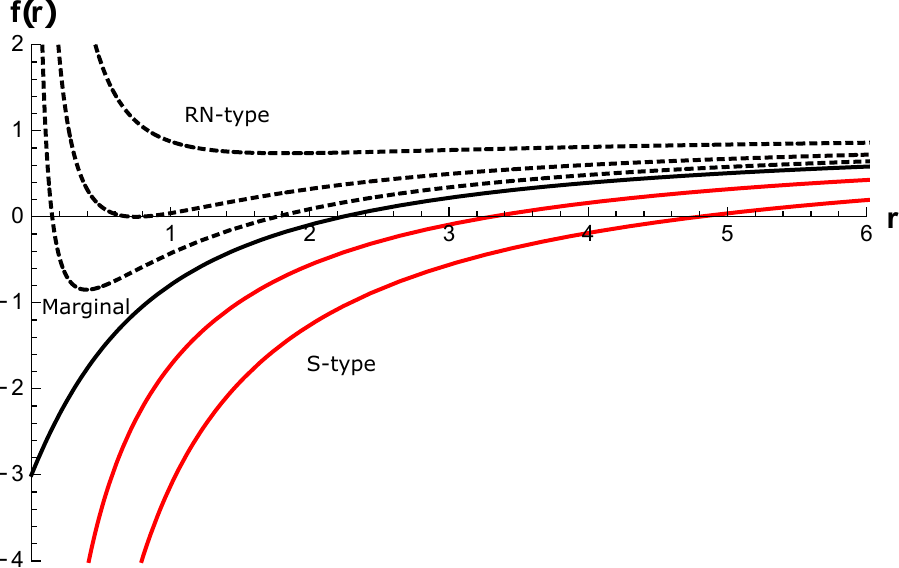}  
\caption{Different types of solutions corresponding to the metric function~(\ref{mf-2}) for different values of $M$. Here $b=4$, $q=1$ and $\Lambda = 0$ are used.}
\label{Fig-mf-2}
\end{figure}

Considering the results~(\ref{a2}) and~(\ref{Mex2}), then as before we can obtain the table~\ref{tab: lagr-2}. Figure~(\ref{Fig-mf-2}) illustrates the different types of solutions that can be classified for the metric function given by Eq.~(\ref{mf-2}). Again the marginal case that we illustrate in this figure corresponds to a one of a regular metric function at the origin. However, it is also possible to obtain a solution belonging to the marginal case that presents a naked singularity.

\begin{table}[h]
\centering
\begin{tabular}{cccc|}                                                                                                                                                                  \\ \hline
\multicolumn{2}{|c|}{Conditions}                                                                                                                                       & \multicolumn{1}{c|}{Type}     &  Horizons \\ \hline
\multicolumn{2}{|c|}{$M > a$}                                                                                                                               & \multicolumn{1}{c|}{S}        & one           \\ \hline
\multicolumn{1}{|c|}{\multirow{3}{*}{$M < a$}} & \multicolumn{1}{c|}{\begin{tabular}[c]{@{}c@{}}$M<M_{ex}$  \\ $b q > 1/2$\end{tabular}}  & \multicolumn{1}{c|}{RN}       & zero          \\ \cline{2-4} 
\multicolumn{1}{|c|}{}                                  & \multicolumn{1}{c|}{\begin{tabular}[c]{@{}c@{}}$M>M_{ex}$\\ $b q > 1/2$\end{tabular}} & \multicolumn{1}{c|}{RN}       & two           \\ \cline{2-4} 
\multicolumn{1}{|c|}{}                                  & \multicolumn{1}{c|}{\begin{tabular}[c]{@{}c@{}}$M=M_{ex}$\\ $b q > 1/2$\end{tabular}}              & \multicolumn{1}{c|}{RN}       & one           \\ \hline
\multicolumn{1}{|c|}{\multirow{2}{*}{$M = a$}}           & \multicolumn{1}{c|}{$b q > 1/2$}                                                                     & \multicolumn{1}{c|}{Marginal} & one           \\ \cline{2-4} 
\multicolumn{1}{|c|}{}                                  & \multicolumn{1}{c|}{$b q < 1/2$}                                                                        & \multicolumn{1}{c|}{Marginal} & zero          \\ \hline
\end{tabular}
\caption{Classification of the types of solutions that can be obtained from the metric function~(\ref{mf-2}) as a function of the mass $M$.}
\label{tab: lagr-2}
\end{table}


\section{Smarr formula and first law of thermodynamics}

Another characteristic of these two Born-Infeld type models is that a Smarr formula can be obtained from the first law of black hole thermodynamics. For this we must introduce a new conjugate thermodynamic quantity to each nonlinearity parameter for each model, as shown below.

\smallskip

In both cases we consider the cosmological constant $\Lambda$ as a thermodynamic pressure according to the relation
\begin{equation}
P = - \frac{\Lambda}{8 \pi}   \, ,
\label{press}
\end{equation}  
where its thermodynamic conjugate is the volume inside the horizon $V=4\pi r_+^3/3$. Likewise, the black hole entropy is $S = \pi r_+^2$ in both cases.

\subsection{Smarr formula for the first model}

From the metric function~(\ref{mf-1}) we can obtain the Hawking temperature on the event horizon by using the surface gravity $\kappa$ as
\begin{align}
T &= \frac{\kappa}{2 \pi}=\frac{1}{4 \pi} \frac{d f \left( r_+\right)}{d r}
\\
T &=\frac{-b q^2-b \Lambda  r_+^4+b r_+^2-\Lambda  q r_+^2+q}{4 \pi  b r_+^3+4 \pi  q r_+} \, .
\label{temp-1}
\end{align}

To calculate the electric potential $\Phi$ on the event horizon we use the expression for the electric field given in Eq.~(\ref{EF-1})
\begin{align}
\begin{split}
\Phi(r_+) = &\int_{r_+}^{\infty} E(r)dr=\frac{1}{8} \Bigg[\frac{2 b q r_+}{b r_+^2+q}-6 \sqrt{b q} 
\tan ^{-1}\Bigg(r_+ +
   \sqrt{\frac{b}{q}}\Bigg)+3 \pi  \sqrt{b q}\Bigg] \, .
\label{pot-1}   
\end{split}
\end{align}

From the metric function evaluated at the horizon, we can obtain the mass of the black hole that depends on the entropy $S$, the pressure $P$, the electric charge $q$ and the nonlinearity parameter $b$
\begin{align}
\begin{split}
M(S,P,q,b) = \frac{1}{12} \Bigg[
&-6 \sqrt{b q^3} \tan ^{-1}\Bigl(\sqrt{\frac{bS}{\pi q}}\Bigl) + 
3 \pi  \sqrt{b q^3}+\frac{2 \sqrt{S} (8 P
   S+3)}{\sqrt{\pi }}\Bigg] \, .
\end{split}
\end{align}
From here, we calculate the black hole temperature

\begin{align}
\begin{split}
T = &\left( \frac{\partial M}{ \partial S}\right)_{q,P,S}
\\
T=& \frac{\pi  q (-b q+8 P S+1)+b S (8 P S+1)}{4 \sqrt{\pi S} (b S+\pi  q)} \, .
\end{split}
\end{align}
Similarly, the electric potential on the event horizon is 
\begin{align}
\begin{split}
\Phi\left(r_+ \right) = &\left( \frac{\partial M}{ \partial q}\right)_{S,P,b}
\\
\Phi\left(r_+ \right) = &\frac{\sqrt{b q} }{8} \Bigg[3 \pi  + \frac{2 \sqrt{\pi bq S }}{bS+\pi  q} -
6 \tan ^{-1}\left(\sqrt{\frac{bS}{\pi q}}\right)\Bigg]\, .
\end{split}
\end{align}
These last two expressions agree with those given in Eqs.~(\ref{temp-1}) and~(\ref{pot-1}) respectively.

\smallskip

We can also get the black hole volume $V$ as the conjugate quantity to pressure $P$
\begin{equation}
V=  \left( \frac{\partial M}{ \partial P}\right)_{S,q,b}\, .
\end{equation}
The differentiation of the mass $M(S, P, q, b)$, leads to the first law
\begin{align}
\begin{split}
d M = & \left( \frac{\partial M}{ \partial S}\right)_{q,P,b} d S +  						\left( \frac{\partial M}{ \partial P}\right)_{S,q,b} d P + 
\\
		& \left( \frac{\partial M}{ \partial q}\right)_{S,P,b} d q + 
		   \left( \frac{\partial M}{ \partial b}\right)_{S,q,P} d b   \, ,
\end{split}
\end{align}
where similarly to Refs. \cite{Gunasekaran:2012dq, Yi-Huan:2010jnv} and have introduced the quantity $B$ as the conjugate quantity to parameter $b$. That is,
\begin{align}
\begin{split}
B =  & \left( \frac{\partial M}{ \partial b}\right)_{S,q,P}
\\
B = & \frac{q^{3/2}}{4\sqrt{b}}
\left[\frac{\pi }{2}-\tan^{-1}\left(\sqrt{\frac{bS}{\pi q}}\right)
   -\frac{\sqrt{\pi b q S }}{b S+\pi  q}\right]\, .    
\end{split}
\end{align}

We can obtain the corresponding Smarr formula by following the scaling arguments in Ref.~\cite{Kastor:2009wy}. The mass M is a homogeneous function of degree 1. If we consider that $L$ has units of length, then from the dimensional analysis we obtain the following scaling relations $S \propto L^2$, $P \propto L^{-2}$, $q \propto L^1$ and $b \propto L^{-1}$. Euler theorem allows us to obtain the following Smarr formula
\begin{align}
\begin{split}
M = & (2)\left ( \frac{ \partial M} { \partial S } \right)   S + (-2) \left ( \frac{ \partial M} { \partial P } \right)  P +
(1)\left ( \frac{ \partial M} { \partial q } \right)   q
+ (-1) \left(\frac{\partial M}{\partial b}\right) b
\,\,\label{smarr-scaling} \,  ,
\end{split}
\end{align}
which we rewrite as
\begin{equation}
M = 2 TS - 2VP + \Phi q - B b 
\,\,\label{smarr-formula} \,  .
\end{equation}

\subsection{Smarr formula for the second model}

In a similar way we can calculate the same thermodynamic quantities as above. The Hawking temperature on the event horizon $r_+$ is obtained using the metric function expression given by Eq.~(\ref{mf-2})
\begin{align}
\begin{split}
T = & \frac{\kappa}{2 \pi}=\frac{1}{4 \pi} \frac{d f \left( r_+\right)}{d r} 
\\
T = &-\frac{1}{4 \pi r_+}\Bigg[ -2 b^{3/2} r_+\sqrt{b r_+^2+2 q} 
+ 2 b^2 r_+^2+2 b q+\Lambda  r_+^2-1 \Bigg]\, .
\label{temp-2}
\end{split}
\end{align}

The electric potential $\Phi$ at the event horizon is calculated with the electric field given by Eq.~(\ref{EF-2})\begin{equation}
\Phi(r_+)= \int_{r_+}^{\infty} E(r)dr=\sqrt{b \left(b r_+^2+2 q\right)}-b r_+ \, .
\label{pot-2}
\end{equation}

From the metric function evaluated at $r = 0$, we now get
\begin{align}
\begin{split}
M(S,P,q,b) = &\frac{2 b^{3/2} S \sqrt{b S+2 \pi  q}-2 b^2 S^{3/2}}{6 \pi ^{3/2}} + 
\frac{\pi  \sqrt{S} (-6 b q+8 P S+3)}{6 \pi ^{3/2}}+
\frac{4 \pi  \sqrt{b} q \sqrt{b S+2 \pi  q}}{6 \pi ^{3/2}} \, .
\end{split}
\end{align}
This function allows us to calculate the following quantities.
The black hole entropy as conjugate to the temperature
\begin{align}
\begin{split}
T = & \left( \frac{\partial M}{ \partial S}\right)_{q,P,S}
\\
T =& \frac{1}{4 \pi^{3/2} \sqrt{S}} 
\Bigg[
2 b^{3/2} \sqrt{S} \sqrt{b S+2 \pi  q} \ -
2 b^2 S-2 \pi  b q+8 \pi  P S+\pi
\Bigg]
\end{split}
\end{align}
The electric potential on the event horizon as conjugate to the electric charge $q$
\begin{equation}
\Phi\left(r_+ \right)=\left( \frac{\partial M}{ \partial q}\right)_{S,P,b}= \frac{\sqrt{b} \sqrt{b S+2 \pi  q}-b \sqrt{S}}{\sqrt{\pi }}\, .
\end{equation}
Note that the same results are obtained in Eqs.~(\ref{temp-2}) and~(\ref{pot-2}) respectively.

\smallskip

Also one can get the black hole volume $V$ as the conjugate quantity to pressure $P$
\begin{equation}
V=  \left( \frac{\partial M}{ \partial P}\right)_{S,q,b}\, .
\end{equation}
And the vacuum polarization $B$ as the conjugate quantity to $b$
\begin{align}
\begin{split}
B =  & \left( \frac{\partial M}{ \partial b}\right)_{S,q,P}
\\
= & \frac{1}{3 \pi ^{3/2} \sqrt{b}}
\Bigg[
-2 b^{3/2} S^{3/2}+2 b S \sqrt{b S+2 \pi  q}
-3 \pi q \sqrt{bS} + \pi  q  \sqrt{b S+2 \pi  q}
\Bigg]
\end{split}
\end{align}

As above, the differentiation of the mass $M(S, P, q, b)$, leads to the first law of thermodynamics
\begin{equation}
d M =T d S+ V d P+ \Phi d q + B d b   \, ,
\end{equation} 
and a consistent Smarr formula
\begin{equation}
M = 2 TS - 2VP + \Phi q - B b 
\,\,\label{smarr-formula-2} \,  .
\end{equation}
In Ref.~\cite{Mazharimousavi:2019sgz} the Smarr formula is different, a cubic term of the electrical potential appears, since the quantity $B$ was not introduced.

\subsection{Heat capacity and thermal stability of the solutions}

In this subsection we discuss the thermal stability of the black hole solutions we have presented. For this purpose we calculate the heat capacity for the values of the parameters $q$, $b$ and $\Lambda$ shown in the figures. Additionally, we study the sign of the temperature where the negative sign indicates that a solution is non-physical.

The heat capacity for each of the models can be obtained with the expression (see for example Ref.~ \cite{Hendi:2018sbe})
\begin{equation}
C_{q,P} = T \left(\frac{\partial S}{\partial T}\right)_{q,P} = T \frac{\left(\frac{\partial S}{\partial r_+}\right)_{q,P}}{\left(\frac{\partial T}{\partial r_+}\right)_{q,P}}
\,\,\label{} \,  .
\end{equation} 

For the first model, we use the temperature given by Eq.~(\ref{temp-1}) and obtain
\begin{equation}
C_{q,P} = \frac{2 \pi  r_+^2 \left(q+b r_+^2\right) \left(b q^2+q \left(r_+^2 \Lambda -1 \right)+b r_+^2 \left(r_+^2 \Lambda -1 \right)\right)}{q^2-b q^3+2 b q r_+^2 - 3 b^2 q^2 r_+^2 + b^2 r_+^4 + r_+^2 \Lambda \left(q + b r_+^2\right)^2}
\,\,\label{C-m1} \,  .
\end{equation} 

In the case of the second model, the temperature given by Eq.~(\ref{temp-2}) allows us to obtain
\begin{equation}
C_{q,P} = \frac{2 \pi  r_+^2 \sqrt{2 q+b r_+^2} \left(1 - 2 b q - 2 b^2 r_+^2 + 2 r_+ \sqrt{b^3 \left(2 q+b r_+^2\right)} - r_+^2 \Lambda \right)}{2 b^{5/2} r_+^3 + 2 b q \sqrt{2 q+b r_+^2}-2 b^2 r_+^2 \sqrt{2 q + b r_+^2}-\sqrt{2 q + b r_+^2} \left(1+r_+^2 \Lambda \right)}
\,\,\label{C-m2} \,  .
\end{equation} 

\begin{figure}[h]
\centering
\includegraphics[scale=0.55]{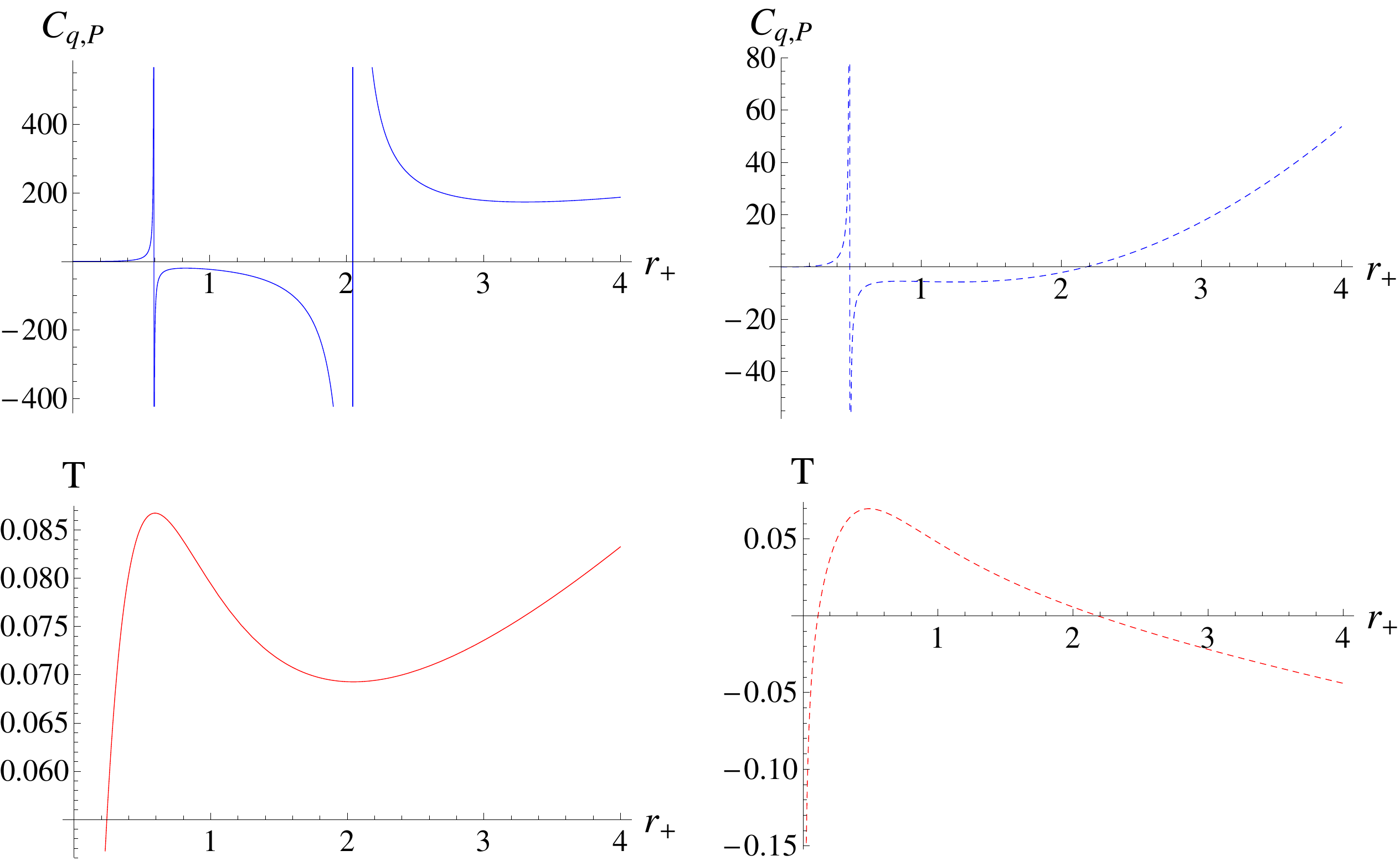}  
\caption{First model. Heat capacity $C_{q,P}$ and temperature T versus $r_{+}$ for $q = 0.5$, $b=2.1$ and $\Lambda = -0.2$ (solid line) and $\Lambda = 0.2$ (dashed line).}
\label{CT-m1}
\end{figure}
The diagrams in Fig.~\ref{CT-m1} correspond to the first model considering the values for the parameters indicated in the caption. In the AdS case (solid line diagrams) the heat capacity presents two divergent points (phase transition points), between which we have an interval of unstable solutions (i.e., $C_{q,P} < 0$). If we enlarge the scale of the horizontal axis we notice that the same situation occurs for $r_+ < 0.1$. In the remaining intervals, stable solutions are found. On the other hand, the temperature is negative only for $r_+ < 0.1$. In summary, for this case we have stable physical solutions in a small interval before the first divergent point and for points from the second divergent point to infinity.
In the dS case (dashed line diagrams) we find stable physical solutions only in the small interval between the single point of divergence and the point where the temperature becomes negative near the origin.
\begin{figure}[h]
\centering
\includegraphics[scale=0.55]{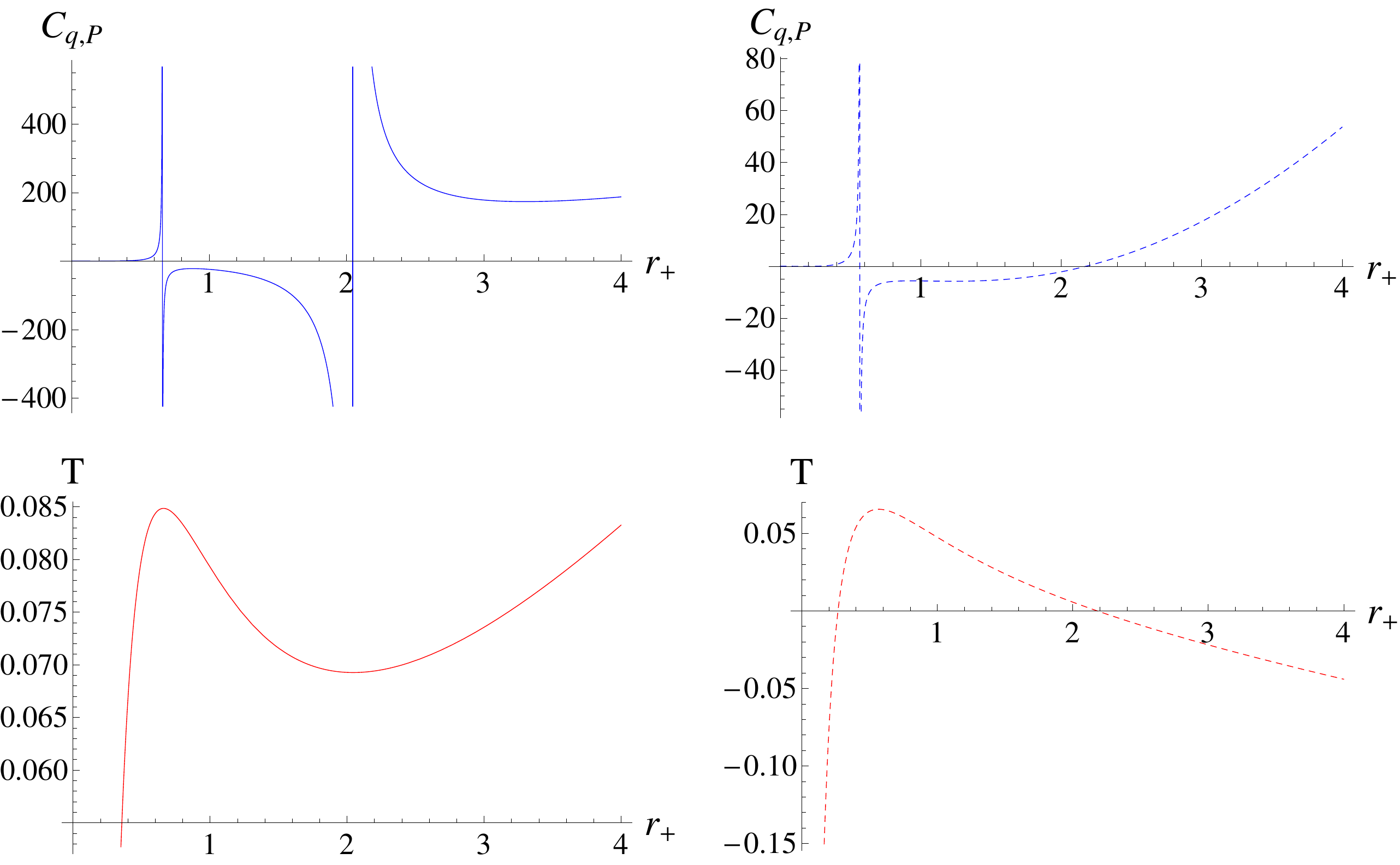}  
\caption{Second model. Heat capacity $C_{q,P}$ and temperature T versus $r_{+}$ for $q = 0.5$, $b=2.1$ and $\Lambda = -0.2$ (solid line) and $\Lambda = 0.2$ (dashed line).}
\label{CT-m2}
\end{figure}
\begin{figure}[h]
\centering
\includegraphics[scale=0.6]{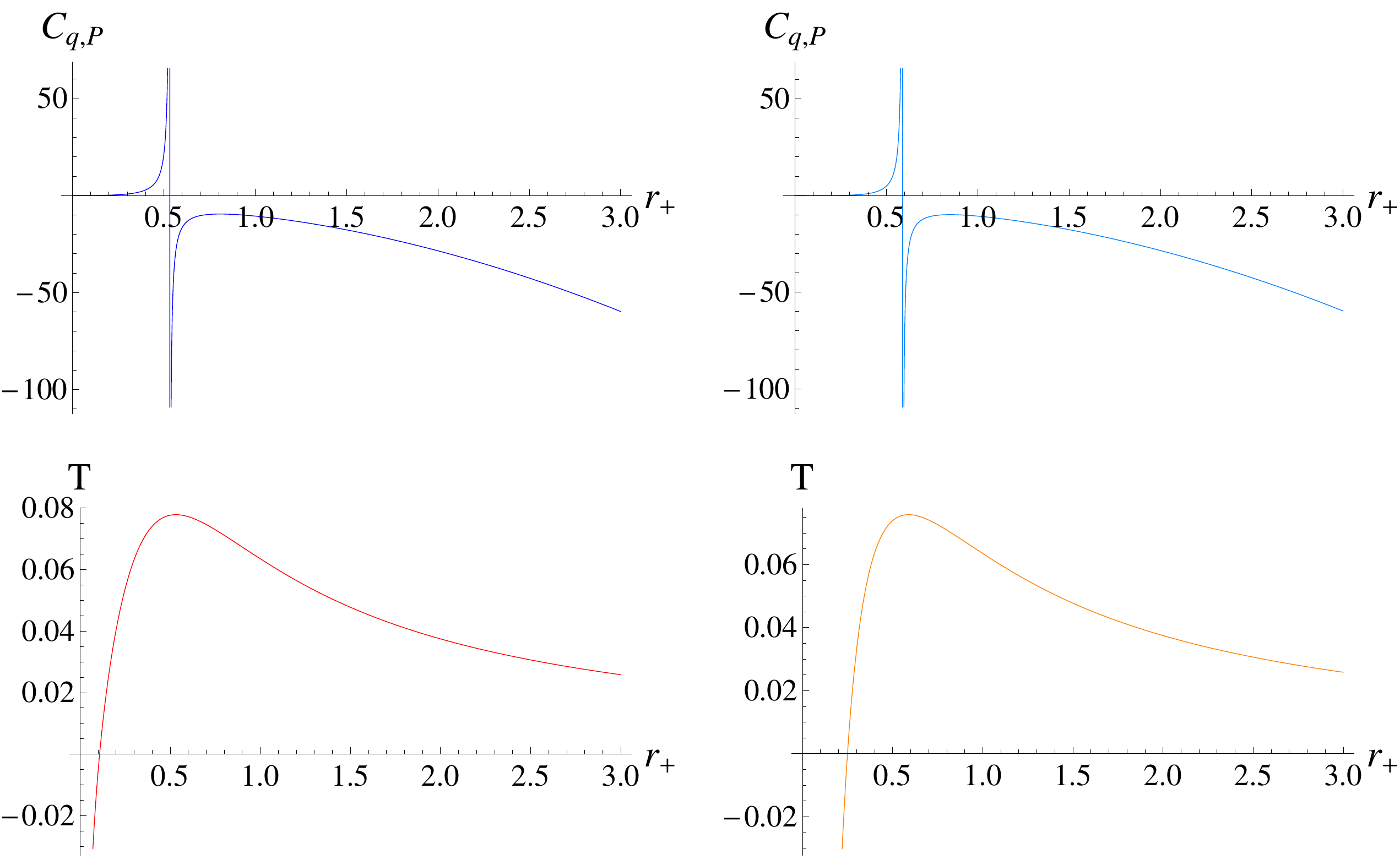}  
\caption{Heat capacity $C_{q,P}$ and temperature T versus $r_{+}$ for $q = 0.5$, $b=2.1$ and $\Lambda = 0$, for the first model (blue and red) and for the second model (light blue and light red).}
\label{0-models}
\end{figure}

If for the second model, we consider the same values for the parameters as for the first model, we find that the graphs behave in a similar way. The numerical differences are minimal, see the diagrams in Fig.~\ref{CT-m2}. Therefore the conclusion is similar as before.

Finally, Fig.~\ref{0-models} illustrates the behavior of heat capacity and temperature considering $\Lambda = 0$  for the two models we are analyzing.
For the same choice of parameter values, we notice that the behavior is similar for both models, that is, in the small interval that is before the only point of divergence, we find stable physical solutions.

\section{Quasinormal spectra}

This section is devoted to introduce essential ingredients to study scalar perturbations in a Born-Infeld-like background. We will focus on a massless scalar field minimally coupled to gravity for simplicity.
%
%
\subsection{Wave equation for scalar perturbations}
%
We start considering the propagation of a test real scalar field, $\Phi$, in a fixed gravitational background in a four-dimensional space-time. Considering $S[g_{\mu \nu} ,\Phi]$ the corresponding action, then we have the following expression 
\begin{align}
S[g_{\mu \nu} ,\Phi] \equiv \frac{1}{2} \int \mathrm{d}^4 x \sqrt{-g}
\Bigl[
\partial^{\mu} \Phi \partial_{\mu} \Phi 
\Bigl]\,.
\end{align}
Utilizing the well-known Klein-Gordon equation (see for instance \cite{crispino,Pappas1,Pappas2,Panotopoulos:2019gtn,Gonzalez:2022ote,Rincon:2020cos} and references therein)
\begin{equation}
\frac{1}{\sqrt{-g}}\partial_{\mu}\left(\sqrt{-g}g^{\mu\nu}\partial_{\nu}\Phi\right) = 0.
\end{equation}
As usual, the Klein-Gordon equation may be solved applied the method of separation of variables in the appropriate coordinate system, taking into consideration the symmetries of the metric tensor. Therefore, we propose as an ansatz for the wave function the following separation of variables in spherical coordinates $r,\theta,\phi$ as follows
\begin{equation}
\Phi(t, r, \theta, \phi) = e^{-i\omega t}\frac{\psi(r)}{r}Y_{\ell m}(\theta, \phi),
\end{equation}
where $Y_{\ell m}(\theta, \phi)$ are the usual spherical harmonics which depend on the angular coordinates only \cite{book}, and
$\omega$ is the unknown frequency to be determined imposing the appropriate boundary conditions. 

%
\begin{align}
\begin{split}
& \frac{\omega^{2}r^{2}}{f(r)} + \frac{r}{\psi(r)}\frac{d}{dr}\left[r^{2}f(r)\frac{d}{dr}\left(\frac{\psi(r)}{r}\right)\right] +
\frac{1}{Y(\Omega)}\left[\frac{1}{\sin\theta}\frac{\partial}{\partial\theta}\left(\sin\theta\frac{\partial Y(\Omega)}{\partial\theta}\right)\right] +
\frac{1}{\sin^{2}\theta}\frac{1}{Y(\Omega)}\frac{\partial^{2}Y(\Omega)}{\partial\phi^{2}} = 0,
\label{KG}
\end{split}
\end{align}
The angular part also satisfies
\begin{align}
    \begin{split}
&\frac{1}{\sin\theta}\frac{\partial}{\partial\theta}\left(\sin\theta\frac{\partial Y(\Omega)}{\partial\theta}\right) + \frac{1}{\sin^{2}\theta}\frac{\partial^{2}Y(\Omega)}{\partial\phi^{2}} = 
-\ell(\ell + 1)Y(\Omega).
\end{split}
\end{align}
with $\ell(\ell + 1)$ being the corresponding eigenvalue, and $\ell$ is the angular degree. 
The Klein-Gordon equation (Eq. (\ref{KG})) is left with the radial part in the tortoise coordinate $r_{*}$:
\begin{equation}
\frac{\mathrm{d}^{2}\psi(r_*)}{\mathrm{d}r_{*}^{2}} + \left[\omega^{2} - V(r_*)\right]\psi(r_*) = 0,
\end{equation}
where we have used the standard tortoise coordinate definition to transform the problem in its Schr{\"o}dinger-like form, i.e., 
\begin{align}
    r_{*}  \equiv  \int \frac{\mathrm{d}r}{f(r)}\, .
\end{align}
while the effective potential barrier, $V(r)$, is computed to be
\begin{equation}
V(r) = f(r)
\Bigg[ 
\frac{\ell(\ell + 1)}{r^{2}} + \frac{f'(r)}{r}
\Bigg] ,\label{poten}
\end{equation}
and the prime denotes a derivative with respect to $r$.

\smallskip

Finally, the wave equation must be supplemented by the following boundary conditions
\begin{equation}
\Phi \rightarrow \: \exp( i \omega r_*), \; \; \; \; \; \; r_* \rightarrow - \infty
\end{equation}
\begin{equation}
\Phi \rightarrow \: \exp(-i \omega r_*), \; \; \; \; \; \; r_* \rightarrow  \infty
\end{equation}

Given the time dependence of the scalar wave function, $\Phi \sim \exp(-i \omega t)$, a frequency with a negative imaginary part implies a decaying (stable) mode, whereas a frequency with a positive imaginary part implies an increasing (unstable) mode.

\smallskip

\begin{figure*}[ht!]
\centering
\includegraphics[scale=0.9]{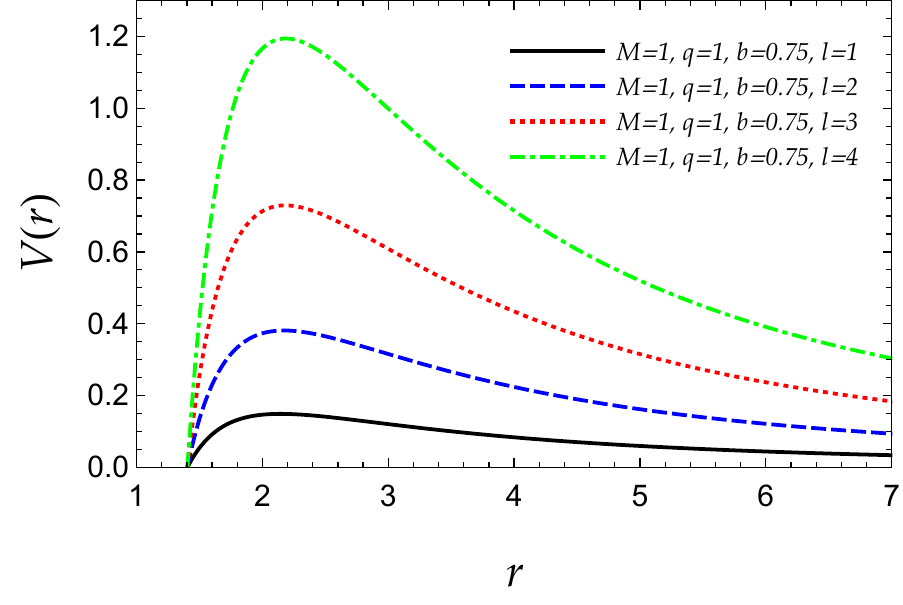} \
\includegraphics[scale=0.9]{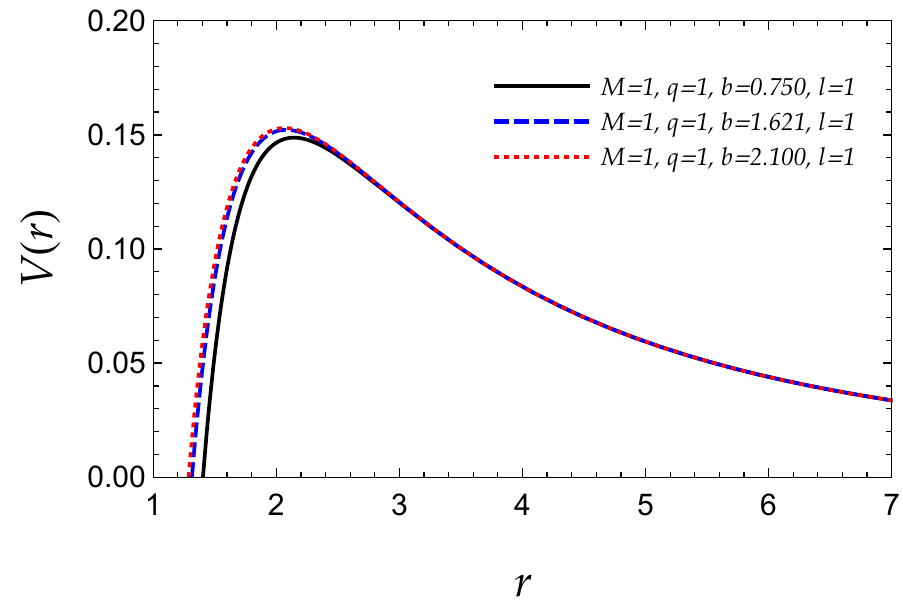} \
\\
\includegraphics[scale=0.9]{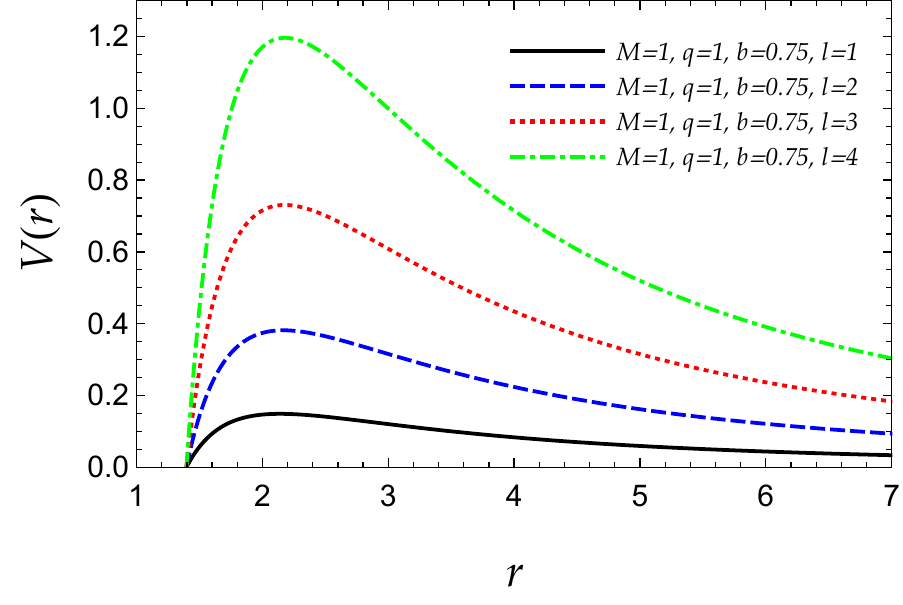} \
\includegraphics[scale=0.9]{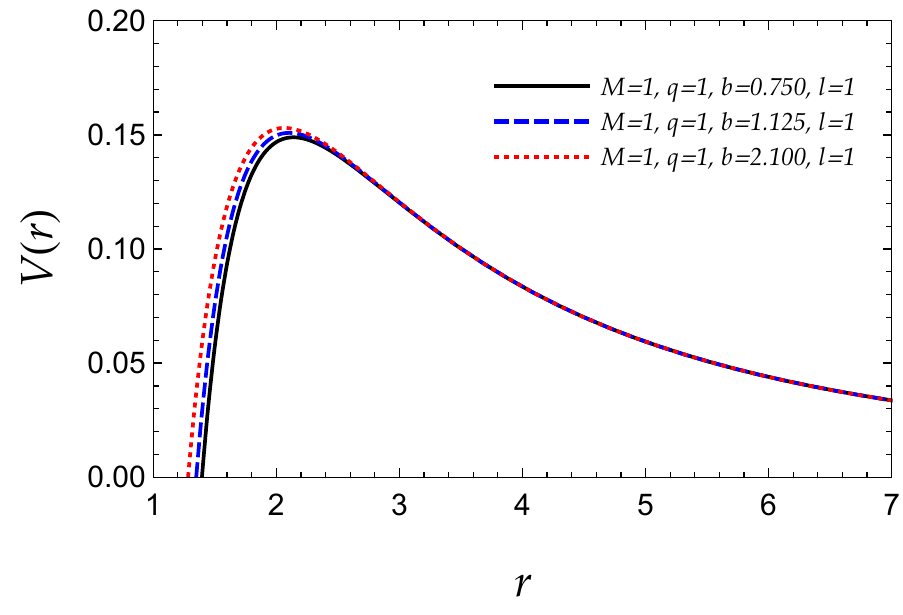} \
\caption{
Effective potential barrier for scalar perturbations against the  radial coordinate for the parameters shown in the panels. The first row corresponds to the first model, whereas the second row shows the behavior of the potencial corresponding to the second model. In all four panels we have assumed $\Lambda = 0$.
}
\label{fig:1} 	
\end{figure*}

In Fig.~\ref{fig:1}, we show the effective potential for the two models discussed in this manuscript. In particular, the first row corresponds to the first model, and the second row to the second model. The qualitative behavior of the effective potential for both cases is essentially the same for the numerical values used. 

\smallskip

The first column corresponds to the effective potential for the first model (top) and the second model (bottom) when the angular degree $\ell$ varies. Thus, we observe that when we increase the angular number $\ell$, the effective potential increases too. Similarly, when $\ell$ increases, $r_{\text{max}}$ is slightly shifted to the right (in both cases).

\smallskip

The second column corresponds to the effective potential for the first model (top) and the second model (bottom) when the non-linear parameter $b$ varies. Similarly to the previous situation,  we observe that when we increase the parameter $b$, the effective potential increases too, but now, the difference in intensity is soft. Finally, 
when $b$ increases, $r_{\text{max}}$ is slightly shifted to the left (in both cases).


\subsection{Numerical computation: WKB method}

Obtaining exact analytic expressions for the quasinormal spectra of black holes is possible only in a few number of cases, for instance:
i) when the effective potential barrier acquires the form of the P{\"o}schl-Teller potential \cite{potential,ferrari,cardoso2,lemos,molina,panotop1}, 
or 
ii) when the corresponding differential equation (for the radial part of the wave function) can be recast into the Gauss' hypergeometric function \cite{exact1,exact2,exact3,Gonzalez:2010vv,exact4,exact5,exact6}. 

\smallskip

In general, it is not possible to obtain an exact analytic solution due to the complexity and non-trivial structure of the differential equation involved. This is the reason why it is necessary to employ one of the numerical schemes available in the literature. 

\smallskip

Up to now there a variety of methods used to obtain, in a very good accuracy, the QN spectra of black holes. One can mention for example the Frobenius method, the generalization of the Frobenius series, fit and interpolation approach, the method of continued fraction, although more methods are known to exist. For more details the interested reader may consult for instance \cite{review3}, and for recent works \cite{Singh:2022ycn,Nomura:2021efi,Sakalli:2022xrb} and references therein.

\smallskip

Given that the WKB semi-classical approximation \cite{wkb0,wkb1,wkb2,wkb3,wkb4} is well-known, we shall avoid the inclusion of unnecessary details. 
Adopting the WKB method, the QN spectra may be computed via the following expression
\begin{equation}
\omega_n^2 = V_0+(-2V_0'')^{1/2} \Lambda(n) - i \nu (-2V_0'')^{1/2} [1+\Omega(n)]\,,
\end{equation}
where 
i) $V_0''$ is the second derivative of the potential evaluated at the maximum, 
ii) $\nu=n+1/2$, $V_0$ is the maximum of the effective potential barrier, 
iii) $n=0,1,2...$ is the overtone number, 
while the functions
$\Lambda(n), \Omega(n)$ are long and complicated expressions of $\nu$ (and derivatives of the potential evaluated at the maximum), and they can be found for instance in \cite{wkb3}. 

\smallskip

At this point it should be mentioned that the 3rd order approximation was first constructed by Iyer and Will in \cite{wkb1}, and it was subsequently extended to higher orders. Thus, to perform our computations, we have used here a Wolfram Mathematica \cite{wolfram} notebook utilizing the WKB method at any order from one to six \cite{code}.

\smallskip

In the present work we have adopted the WKB method at sixth order. Besides, it should be mentioned that for a given angular degree, $\ell$, we have considered values $n < \ell$ only, since it is known that the method works well for high angular degrees. For higher order WKB corrections (and recipes for simple, quick, efficient and accurate computations) see \cite{Opala,Konoplya:2019hlu,RefExtra2}. In particular, we should mention that as the WKB series converges only asymptotically, there is no mathematically strict criterium for evaluation of an error according to \cite{Konoplya:2019hlu}. However, the sixth/seventh order usually produce the best results. In that direction, taking into account the Pad{\'e} approximations we can have a higher accuracy of the WKB approach. This analysis, however, will be performed in a future study.

\begin{table*}
\centering
\caption{Quasinormal frequencies (varying $\ell$ and $n$) for the two models and the three behaviors (S-type, RN-type and Marginal) considered in this work. The numerical values of the parameters $M,q,b$
assumed to obtain the QNMs are the ones shown in Fig.~\ref{fig:1}.}
\begin{tabular}{c | c | c}
  & \multicolumn{2}{c}{{\sc QN frequencies for Scalar perturbations}} \\
  & First model & Second model  \\
\hline
\hline
Type S $(\ell=2)$  	 &  {$\omega(n=0)$ = 0.605941 - 0.101944 I}       & {$\omega(n=0)$ = 0.621496 - 0.0990191 I}  \\
						  	 & {$\omega(n=1)$ = 0.592259 - 0.309586 I}       & {$\omega(n=1)$ = 0.749904 - 0.2436500 I}  \\
\hline
Type RN $(\ell=2)$   & {$\omega(n=0)$ = 0.618099 - 0.096235 I}       & {$\omega(n=0)$ = 2763.360 + 0.2359470 I}  \\
               				 & {$\omega(n=1)$ = 0.605848 - 0.291446 I}       & {$\omega(n=1)$ = 9962.160 + 0.7206980 I}  \\
\hline
Type RN $(\ell=12)$  & {$\omega(n=0)$ = 3.07474 - 0.0958944 I}       & {$\omega(n=0)$ = 40.4016 - 0.00272247 I}  \\ 
           				     & {$\omega(n=1)$ = 3.07214 - 0.2877990 I}       & {$\omega(n=1)$ = 145.237 + 0.00786804 I} \\
\hline
 Type RN $(\ell=20)$  & {$\omega(n=0)$ = 5.04189 - 0.095885 I}        & {$\omega(n=0)$ = 0.077788 - 5.42437 I} \\
              				  & {$\omega(n=1)$ = 5.04030 - 0.287699 I}        & {$\omega(n=1)$ = 0.029747 - 26.2024 I} \\  
               				 & {$\omega(n=2)$ = 5.03713 - 0.479642 I}        & {$\omega(n=2)$ = 0.019038 + 68.0079 I} \\
\hline
Type M $(\ell=2)$    & {$\omega(n=0)$ = 0.615788 - 0.097606 I}       & {$\omega(n=0)$ = 0.937076 - 0.0634837 I} \\  
                				& {$\omega(n=1)$ = 0.599094 - 0.302352 I}       & {$\omega(n=1)$ = 2.594380 - 0.0642422 I} \\
\end{tabular}
\label{table:Third_set}
\end{table*}

\smallskip

Our findings may be summarized as follows: Regarding the sign of the imaginary part of the frequencies, in the first model all frequencies are characterized by negative imaginary part, whereas in the second model, the same holds for S-type and Marginal behaviour. In contrary, in the RN-type behaviour, for $\ell=2$ both modes are characterized by a positive imaginary part, and then considering higher values of the angular degree, after a certain value of the overtone, $n$, the imaginary part changes sign. 

\smallskip

Regarding the impact of $n,\ell$ on the spectra, a direct comparison between the two models reveals the following features: a) Regarding the S-type behaviour, the imaginary part becomes more negative with $n$ in both models, while the real part decreases with $n$ in the first model and increases with $n$ in the second model. b) Regarding the Marginal behaviour, just like in the previous case, the imaginary part becomes more negative with $n$, although only slightly in the second model, while the real part decreases with $n$ in the first model and increases with $n$ in the second model. 
Finally, in the RN-type behaviour, the real part of the frequencies increases with $\ell$ in the first model, and
decreases with $\ell$ in the second model.

\smallskip

Before we conclude, and as far as future work is concerned, we add here as a final comment that it would be interesting to investigate even further the properties of the novel black hole solutions discussed in the present article. In particular, it would be worth studying the QN spectra for Dirac and Maxwell fields, and also the deflection of light as well as their shadow casts \cite{cunha}. We hope to be able to address some of those exciting topics in forthcoming publications.

\section{Conclusions}

In the present paper, we investigated in detail two novel electrically charged black hole solutions within non-linear electrodynamics of
Born-Infeld type in four-dimensional space-time. In particular, we computed the metric function for each Lagrangian density considered here, and we demonstrated that each solution exhibits three distinct behaviours (S-type, RN-type and Marginal) depending on the numerical values of the parameters. Next, we discussed black hole thermodynamics for both models as well as the Smarr formula, which was shown to be compatible with the first law of black hole mechanics. We extended the usual Smarr formula defining a new conjugate thermodynamic quantity associated to the non-linear parameter, $b$.
Finally, we computed the frequencies of some quasinormal modes for scalar perturbations analyzing the propagation of a massless, canonical scalar field varying the parameters $\{b, \ell, n \}$ for certain values of $M,q$. The WKB method of sixth order was employed to compute the modes numerically. We found that as far as the first model is concerned, the frequencies are always characterized by a negative imaginary part, irrespectively of the type of behaviour. Regarding the second model, however, our numerical results show that in the S-type and marginal type, the frequencies are characterized by a negative imaginary part, but in the RN-type some frequencies are characterized by a positive imaginary part. In summary, naively one might expect to see almost identical spectra. Our findings, however, show that although the models look very similar, their quasinormal spectra are characterized by certain differences.

\section*{Acknowledgments}
A.~R. is funded by the María Zambrano contract ZAMBRANO 21-25 (Spain)
(with funding from NextGenerationEU). 
%
S.B-H. acknowledges financial support from Universidad de la Frontera. L.~B. is supported by DIUFRO through the project: DI22-0026.


\section*{Data Availability Statement}

Authors' comment: This is a theoretical study, and for that reason no experimental data are presented. 





\end{document}